\documentclass[twoside,reqno]{qts-proc}
\usepackage{epsfig,cite}
\usepackage{amssymb,amsmath}
\usepackage{times}
\begin{document}

\setlength{\parindent}{15pt} \setlength{\textwidth}{11.2cm}
\setlength{\textheight}{46\baselineskip}
\setlength{\oddsidemargin}{0in} \setlength{\evensidemargin}{0in}
\pagestyle{headings}

\sloppy \raggedbottom \setcounter{page}{1}

\newpage
\setcounter{figure}{0} \setcounter{equation}{0}
\setcounter{footnote}{0} \setcounter{table}{0}
\setcounter{section}{0}

\newcommand{\R}{\mathbb{R}}
\newcommand{\C}{\mathbb{C}}
\newcommand{\Z}{\mathbb{Z}}
\newcommand{\Hb}{\mathbb{H}}

\newcommand{\rSU}{\mathrm{SU}}
\newcommand{\rU}{\mathrm{U}}
\newcommand{\rSO}{\mathrm{SO}}

\newcommand{\rspan}{\mathrm{span}}
\newcommand{\rsolv}{\mathrm{solv}}
\newcommand{\rtr}{\mathrm{tr}}
\newcommand{\rEnd}{\mathrm{End}}

\newcommand{\fso}{\mathfrak{so}}
\newcommand{\fsu}{\mathfrak{su}}
\newcommand{\fg}{\mathfrak{g}}
\newcommand{\fh}{\mathfrak{h}}
\newcommand{\fp}{\mathfrak{p}}
\newcommand{\fk}{\mathfrak{k}}
\newcommand{\fs}{\mathfrak{s}}
\newcommand{\fid}{\mathfrak{i}}
\newcommand{\ft}{\mathfrak{t}}
\newcommand{\fn}{\mathfrak{n}}
\newcommand{\fa}{\mathfrak{a}}
\newcommand{\fsolv}{\mathfrak{solv}}

\newcommand{\cM}{\mathcal{M}}
\newcommand{\cI}{\mathcal{I}}
\newcommand{\cL}{\mathcal{L}}
\newcommand{\cK}{\mathcal{K}}
\newcommand{\cF}{\mathcal{F}}
\newcommand{\cN}{\mathcal{N}}
\newcommand{\cH}{\mathcal{H}}

\newcommand{\e}{\epsilon}

\newcommand{\id}{\relax{\rm 1\kern-.35em 1}}


\rightline{CERN-PH-TH/2005-231}

\rightline{IFIC/05-61}

\rightline{FTUV-05/1122}
\title{Contractions of sigma models and integration of
massive modes}

\runningheads{Mar\'{\i}a A. Lled\'o, O. Maci\'a}{Contractions of
Sigma Models}

\begin{start}


\author{Mar\'{\i}a A. Lled\'o}{1},
\coauthor{Oscar Maci\'a}{1,2},

\address{Address}{1} Departamento de F\'{\i}sica Te\'orica, Universidad de
Valencia and IFIC. C/ Dr. Moliner, 50. 46.100, Burjassot,
Valencia. Spain

\address{Address}{2} Department of Physics, Theory Division CERN, CH
1211 Geneva 23, Switzerland


\begin{Abstract}
We show how the integration of massive modes after a spontaneous
symmetry breaking in a sigma model can often be interpreted as a
contraction, induced by a group contraction, of the target space of
the sigma model.
\end{Abstract}
\end{start}

\section{Introduction}
In this talk we summarize the results of Ref. \cite{aflm}. The main
idea is to find a geometrical way of describing the integration of
massive modes after a gauging of translational isometries. We
consider examples of sigma models that are maximally symmetric
spaces and so they have a solvable Lie group structure. We apply the
theory of contractions of Lie algebras and groups to define
deformations and contractions of the metric of the sigma model, and
we find out for several examples that the geometrical interpretation
of the integration is a generalized contraction.

In section \ref{contractions} we make a review of the theory of
contractions and explain how do we apply it to our examples. In
section \ref{first} we start with a simple example, and we see
that an exact integration or truncation of a theory is related
with a contracted Lie algebra that is isomorphic to the non
contracted one. Then we describe a more complicated model where an
In\"on\"u-Wigner \cite{iw} contraction models the integration of
massive modes after the gauge symmetry breaking . In section
\ref{sugra} we describe a model where instead a generalized or
Weimar-Woods contraction is needed to model the integration. This
last model correspond to a certain supregravity theory.

\section{\label{contractions}Contractions of Lie algebras, groups and symmetric
spaces}

Let {$\fg$} be a finite dimensional Lie algebra with commutator
{$[\,,\,]$}, and {$\e$} a real parameter.

Let $\phi_\e:\fg\rightarrow \fg$ denote a family of linear maps
parametrized by $\e$, such that they are non degenerate except
possibly for {$\e=0$}.

The {\it deformed commutator} on the vector space $\fg$ {$$[X,Y]_\e=
\phi^{-1}_\e([\phi_\e(X),\phi_\e(Y)]), \qquad X,Y\in \fg$$} defines
a Lie algebra {$\fg_\e\approx\fg$} except possibly for {$\e=0$}.
Because of the non degeneracy of $\phi_\e$ at $\e\neq 0$, the
deformed Lie algebra defined by $[\,,\,]_\e$ is isomorphic to the
one defined by $[\,,\,]$.

If the limit {$[X,Y]_c=\lim_{\e\rightarrow 0}[X,Y]_\e$} exists, then
it defines a Lie algebra structure on the vector space {$\fg$}
denoted by {$\fg_c$}. It is a {\it contraction} of the Lie algebra
{$\fg$}. In general, {$\fg$} is not isomorphic to {$\fg_c$}.

These are {\it generalized In\"on\"u-Wigner} or {\it Weimar-Woods}
contractions. The conditions on $\phi_\e$ to have a well defined
bracket $[\,,\,]_c$  are studied in detail in Ref. \cite{iw,ww}.

\bigskip

The standard In\"on\"u-Wigner contraction is obtained when {$\fg$}
is split as {$\fg=\fg_1+\fg_2$}, with {$\fg_1$} a subalgebra, and
{$$\phi_\e=\begin{pmatrix}\id&0\\0&\epsilon\id\end{pmatrix}.$$} The
result has the form {$\fg_c=\fg_1\ltimes \R^n$}.

\bigskip

We can also contract a representation of a Lie algebra to a
representation of the contracted Lie algebra: if {$W$} is a module
for {$\fg$} ({$R:\fg\rightarrow \rEnd{W}$}), one has to find a
linear map {$\psi_\e:W\rightarrow W$} such that the limit
{$$R_c(X)=\lim_{\e\rightarrow 0}\psi_\e^{-1}\circ
R(\phi_e(X))\circ\psi_\e$$} exists. Then we have a representation of
{$\fg_c$}.

For example, for the standard In\"on\"u-Wigner contraction, it is
enough to split {$W=W_1+W_2$} where {$W_1$} is a representation of
{$\fg_1$} and
{$$\psi_\e=\begin{pmatrix}\id&0\\0&\epsilon\id\end{pmatrix}.$$} Then
$R_c$ is well defined and it is a representation of $\fg_c$.

For example, the adjoint representation always admits a contraction
to the adjoint representation of the contracted algebra using {$
\psi_\e=\phi_e$}.

\bigskip

We will consider deformations and contractions of Lie groups and
symmetric spaces via exponentiation of the corresponding Lie
algebras. This will be enough for our purposes, since the models
that we will study admit global exponential coordinates.

\subsection{Symmetric spaces and metrics \label{symmetric spaces}} Let
{$G$} be a semisimple Lie group with Lie algebra {$\fg$} and
consider a {Cartan decomposition} {$\fg=\fh+\fp$}. Let {$H$} be the
maximal compact subgroup of $G$; it has Lie algebra {$\fh$}.

We consider the principal bundle {$G\rightarrow G/H$} and a local
section (or {\it coset representative}) on it
{$L:U\subset_{_{\mathrm{open}}} G/H\rightarrow G$}.

The pull back of the {\it Maurer-Cartan form} on
{$U\subset_{_{\mathrm{open}}}G/H$} decomposes as
{$$L^{-1}dL=(L^{-1}dL)_\fh+(L^{-1}dL)_\fp.$$} Then the invariant
metric on {$G/H$} is
{$$\langle(L^{-1}dL)_\fp,(L^{-1}dL)_\fp\rangle.$$}
{$\langle\;,\;\rangle$} is the {\it Cartan-Killing form} of {$\fg$}.
(One can see for example Ref. \cite{he} for the details).

 By using a representation of {$\fg_\e$} and its exponential, we
can compute the coset representative {$L_\e$} and an
{$\e$}-dependent metric, that is, a deformation of the metric. The
inner product {$\langle\;,\;\rangle$} is the Cartan-Killing form of
{$\fg$},
 so it is independent
of {$\e$}. This implies that the metrics are not isometric. There is
no change of coordinates taking one metric into the other, and the
limit {$\e\rightarrow 0$} produces a non degenerate metric. We can
have Einsten spaces that are deformed to non Einstein metrics, which
shows that the deformation is non trivial, although the Lie algebras
are isomorphic.

Using a Cartan Killing form depending on {$\e$}  would not give a
true deformation of the metric. Moreover, the limit {$\e\rightarrow
0$} will be degenerate, since the group becomes non semisimple.

\bigskip

The {\it Iwasawa } or {\it $KAN$ decomposition} (here $K=H$, $N$ is
nilpotent and $A$ abelian) assures us that the tangent space to the
coset at the identity, {$\fp$} has a solvable Lie algebra structure.
In fact,  {$G/H\sim AN$}  is itself a solvable Lie group inside $G$.
We can consider contractions of this group instead of the whole
group {$G$}, which gives us more possibilities.

\section{\label{first}First examples.}

\paragraph{Exact truncation.} We consider the symmetric space
{$\rSO(1,1+n)/\rSO(1+n)$}, whose solvable Lie algebra is given by
the commutation relations
 {$$[H,Y_a]=Y_a,\quad a=1,\dots n.$$}
 We choose the following coset representative {$L=e^{u^aY_a}e^{\varphi
 H}$}, which
 is an element of the solvable group. The coordinates $u^a, \varphi$ are
  {\it global coordinates} (for a proof see appendix in Ref. \cite{aflm}). Using the
  technic explained in Section \ref{symmetric spaces}, one can compute the metric in these coordinates:
  {$$ds^2=d\varphi^2+e^{-2\varphi }\sum_{a=1}^n(du^a)^2.$$}
  It is immediate to see that the transformations $u^a\rightarrow u^a+c^a$, with $c^a
  \in \R$ are isometries of the metric, generated by $Y^a$. They are
 {\it  translational isometries}.

Suppose that we have a sigma model with such metric. Let us {\it
gauge} one of the translational isometries, say {$Y_n$}. This means
that we introduce a gauge field {$A=A_\mu dx^\mu$} and substitute
{$du^n$} by the covariant differential {$Du^n=du^n +gA$}. We  can
equally use the gauge transformed connection {$\hat A=A+\frac 1g
du^n$}. {Substituting this in the metric we obtain}
{$$ds^2=d\varphi^2+ e^{-2\varphi }\sum_{a=1}^{n-1}(du^a)^2
+e^{-2\varphi }g^2\hat A^2.$$} We observe that the coordinates
{$u^n$} have disappeared from the kinetic term, while the gauge
vector has acquired mass. Moreover, {$\hat A$}
 is decoupled (except
for the factor {$e^{-2\varphi}$}) and the condition {$\hat A=0$} is
consistent with the equations of motion. We can {\it excatly
integrate} (by integrating we mean substituting the equations of
motion) the massive mode, or, in other words, we have a {\it
consistent truncation} of the theory. The remaining massless modes
complete the sigma model with target space {$\rSO(1,n)/\rSO(n)$}.

\bigskip

From the geometric point of view, this is a special case. The
contraction of {$\fsolv(\rSO(1,1+n)/\rSO(1+n))$} with respect to
{$\fsolv(\rSO(1,n)/\rSO(n))$,} does not change the algebra, so there
is no a true contraction. No limit {$\e\rightarrow 0$}, (related
with taking the mass very big)
  has been necessary.

  \paragraph{In\"on\"u-Wigner contraction.}
We consider now the sigma model with target space
{$\rSU(1,1+n)/\rU(1+n)$.} The solvable Lie algebra is
{$$[H,Y_a]=Y_a, \quad [H,Z_a]=Z_a,\quad [Z_a,Y_b]=\delta_{ab}S,$$}
with $a=1,\dots n$.

We will take $n=1$ for clarity. Using global solvable coordinates $
( s, z, y, \phi)$, with coset representative
$$
L( s, z, y, \phi)=\exp{(sS+zZ)}\exp{yY}\exp{\phi H},$$
   we obtain the metric
   {\begin{eqnarray*} ds^2= 2d\phi^2
+\frac{1}{2}e^{-4\phi}ds ds + e^{-4\phi}y d sdz
+\\\frac{1}{2}e^{-4\phi}(e^{2\phi}  +y^2)dz^2
+\frac{1}{2}e^{-2\phi}dy^2,\end{eqnarray*}} that makes manifest the
translational isometries ({$Z, S$}).  In this case we gauge the two
translational isometries by introducing two gauge fields as before.
The modes {$z, s$} are absorbed to give mass to the vectors, but now
other interactions are present. Assuming that the mass of the vector
fields is very big (the kinetic term is very small), we obtain
algebraic equations for the gauge fields.  After the elimination of
the gauge fields using the approximate equations of motion,  we
obtain that the remaining modes {$\phi, y$}) parametrize the
manifold {$\rSO(1,2)/\rSO(2)$}

\bigskip

The same result can be obtained by performing a contraction of
{$\fsolv(\rSU(1,2)/\rSU(2))$} with respect to
{$\fsolv(\rSO(1,2)/\rSO(2))$}. The result is
{$\fsolv(\rSO(1,2)/\rSO(2))\ltimes\R^2$}, with metric {$$ ds^2=
\bigl(2d\phi^2 +\frac{1}{2}e^{-2\phi}dy^2\bigr)
+\frac{1}{2}e^{-4\phi}ds^2 + e^{-2\phi}dz^2.$$} The terms inside the
parenthesis reproduce the metric of {$\rSO(1,2)/\rSO(2)$}. The other
modes
 appear almost decoupled, in a way that it is consistent to truncate them to {$z, s=0$}.

 A we will see in next section, for more involved metrics it is not enough with a standard In\"on\"u-Wigner
  contraction to obtain the simple structure of the above metric. Instead a generalized contraction may be  needed.

\bigskip

Summarizing, we can say that the integration of the
 massive modes can be modeled by a group contraction
 followed by a quotient by an abelian invariant subgroup (or exact truncation).

 \

\section{\label{sugra}A generalized contraction: application to Supergravity}

We consider now the coset space  {$\rU(2,1+n)/(\rU(2)\times
\rU(1+n))$}, with solvable Lie algebra {\begin{eqnarray*}
&&[Z^{ia},Z^{jb}]=\epsilon^{ij}\delta^{ab}T^{(2,0)}\\
&&[Y^{ia},Y^{jb}]=\epsilon^{ij}\delta^{ab}T^{(0,2)}\\
&&[Z^{ia},Y^{jb}]=\delta^{ab}(\delta^{ij}S_2^{(1,1)}+\epsilon^{ij}S_1^{(1,1)})\\
&&[Y^{ia}, S_1^{(1,-1)}]=Z^{ia}\\
&&[Y^{ia}, S_2^{(1,-1)}]=\epsilon^{ij}Z^{ja}\\
&&[T^{(0,2)}, S_\alpha^{(1,-1)}]=2S_\alpha^{(1,1)}\\
&&[S_\alpha^{(1,1)},S_\beta^{(1,-1)}]=\delta_{\alpha\beta}T^{(2,0)}\\
&&[H_+, Z^{ia}]=Z^{ia}\\
&&[H_-, Y^{ia}]=Y^{ia},
\end{eqnarray*}}
with {$i=1,2,\quad \alpha=1,2,\quad a=1,\dots n.$} The superindices
indicate the weights with respect to the Cartan generators.

We consider the coordinates and coset representative
   \begin{equation}
   L(t,\tilde t, \tilde s_\alpha, s_\alpha, z_{ia}, y_{ia},
   \psi,\phi)=A(t,\tilde t, \tilde s_\alpha,
   z_{1a})B(s_\alpha, z_{2a},
   y_{ia})C(\psi,\phi)\label{solvpar}\end{equation} where
   \begin{eqnarray*}A&=&\exp{(tT^{(2,0)}+\tilde t T^{(0,2)}+\tilde
   s_\alpha S^{(1,1)}_\alpha+z_{1a} Z^{1a})}\\
   B&=& \exp{(s_1S_1^{(1,-1)})}\exp{(s_2S_2^{(1,-1)})}
   \exp({z_{2a}Z^{2a}})\exp({y_{2a}Y^{2a}})\exp({y_{1a}Y^{1a}})\\
   C&=& \exp{( \psi H_+ + \phi H_-)}.
   \end{eqnarray*}
   The metric is
 \footnote{This has been computed using the program Wolfram
Research, Inc., Mathematica, Version 5.1, Champaign, IL (2004).}
(sum over repeated indices is understood, and we have used the
short-hand notation $y_1^2=y_{1a}y_{1a}$):

\begin{eqnarray*}
   &&ds^2=
   d\phi^2 +
   d\psi^2+
   e^{-4\psi}dtdt+
   2e^{-4\psi} s_1 dtd\tilde s_1 +
   2 e^{-4\psi} s_2 dtd\tilde s_2\nonumber\\
   && +2e^{-4\psi}z_{2a} dtdz_{1a} +
   2e^{-4\psi}(s_2^2 + s_1^2)dtd\tilde t+
   \frac{1}{2}(e^{-2(\psi+\phi)}+ 2e^{-4\psi}s_1^2)d\tilde s_1 d\tilde s_1
   \nonumber\\
   && +2 e^{-4\psi} s_2  s_1 d\tilde s_1d\tilde s_2  +
   \frac{1}{2} e^{-2(\psi+\phi)}(y_{1}^2+y_{2}^2)d\tilde s_1d s_2+
   e^{-2(\psi+\phi)}y_{1a}y_{2a} d\tilde s_1d s_1 \nonumber\\
   && -e^{-2(\psi+\phi)}y_{1a} d\tilde s_1 dz_{2a}+
   (2e^{-4\psi} s_1 z_{2a} +e^{-2(\psi +\phi)}y_{2a} ) d\tilde
   s_1dz_{1a}\nonumber\\
   && + 2 e^{-4\psi} s_1 (e^{2(\psi-\phi)}+ s_2^2 + s_1^2 ) d\tilde
   s_1d\tilde t +
 \frac{1}{2}(e^{-2(\psi+\phi)}+2e^{-4\psi} s_2^2)d\tilde s_2 d\tilde
   s_2\nonumber\\
   &&+e^{-2(\psi+\phi)}y_{1a}y_{2a} d\tilde s_2 d s_2
   -\frac{1}{2}e^{-2(\psi+\phi)}(y_{1}^2 +y_{2}^2)d\tilde s_2 d s_1\nonumber\\
   &&+e^{-2(\psi+\phi)}y_{2a}d\tilde s_2dz_{2a}+
   (2e^{-4\psi}s_2 z_{2a}+e^{-2(\psi+\phi)}y_{1a})d\tilde s_2dz_{1a} \nonumber\\
   && + 2e^{-4\psi}s_2( e^{2(\psi-\phi)}+ s_2^2 +  s_1^2 )d\tilde s_2
   d\tilde t\nonumber\\
   && +
   \frac{1}{8}e^{-2(\psi+\phi)}(4e^{4\phi}+ 4e^{2\phi}
   (y_{1}^2+y_{2}^2)+ 4(y_{1a}y_{2a})^2+(y_{1}^2+y_{2}^2)(y_{1}^2+y_{2}^2) )d s_\alpha d s_\alpha \nonumber\\
   && - \frac{1}{2}e^{-2(\psi+\phi)}(2e^{2\phi}y_{1b}+(-2(y_{1a}y_{2a})y_{2b}+(y_{1}^2
   +y_{2}^2)y_{1b}))d s_2
   dz_{2b}\nonumber\\&&+
   \frac{1}{2}e^{-2(\psi+\phi)}(2e^{2\phi}y_{2b}+(2(y_{1a}y_{2a})y_{1b}+
   (y_{1}^2+y_{2}^2)y_{2b}))d s_2
   dz_{1b}\nonumber\\
   && + e^{-2(\psi+\phi)}(y_{2}^2 s_1+2y_{2a}  s_2
   y_{1a}+ s_1 y_{1}^2)d s_2 d\tilde t\nonumber\\
    && - \frac{1}{2}e^{-2(\psi+\phi)}(2e^{2\phi}y_{2b}+(2(y_{1a}y_{2a})y_{1b}+(y_{1}^2+y_{2}^2)y_{2b}))
    d s_1dz_{2b}\nonumber \\&&
   - \frac{1}{2}e^{-2(\psi+\phi)}(2e^{2\phi}y_{1b}+(-2(y_{1a}y_{2a})y_{2b}+
   (y_{1}^2+y_{2}^2)y_{1b}))d s_1 dz_{1b}
   \nonumber\\
   &&-e^{-2(\psi+\phi)}(y_{1}^2  s_2 - 2y_{1a} s_1 y_{2a} +  s_2 y_{2}^2 )
   ds_1d\tilde t\nonumber\\&&-
\frac{1}{2}e^{-2(\phi+\psi)}\epsilon_{ij}\epsilon_{mn}(y_{ia}y_{jb})dz_{ma}dz_{nb}\nonumber\\&&
   +\frac{1}{2}e^{-2(\psi+\phi)}(e^{2\phi}\delta_{ab}+(y_{1a}y_{1b}+y_{2a}y_{2b}))dz_{2a}dz_{2b}\nonumber\\
 && -e^{-2(\psi+\phi)}(2y_1 s_1 - 2  s_2 y_2) dz_2
   d\tilde t\nonumber\\&&+
   \frac{1}{2}e^{-4\psi}(e^{2\psi}\delta_{ab}+2z_{2a}z_{2b}+e^{2(\psi-\phi)}
   (y_{1a}y_{1b}+y_{2a}y_{2b}))dz_{1a} dz_{1b}\nonumber\\
   && +(2e^{-4\psi}( s_2^2+ s_1^2)z_{2a} +2e^{-2(\psi+\phi)}(y_{1a} s_2+ s_1
   y_{2a}))dz_{1a}d\tilde t\nonumber\\
   &&+ e^{-4(\psi+\phi)}(e^{2\psi}+e^{2\phi}( s_1^2 + s_2^2)
   )^2 d\tilde t d\tilde t- 2 e^{-4\phi}y_{1a} d\tilde tdy_{2a}\nonumber\\
   && + \frac{1}{2}e^{-4\phi}(e^{2\phi}\delta_{ab}+2y_{1a}y_{1b})dy_{2a} dy_{2b} +
   \frac{1}{2}e^{-2\phi}dy_{1a}dy_{1a} \label{bigmetrice=1}
   \end{eqnarray*}
The generators   {$\{T^{(2,0)},T^{(0,2)},S_\alpha^{(1,1)},Z^{1a}\}$}
are translational isometries.

\vfill\eject

We have the following chain of symmetric spaces and their
corresponding solvable Lie algebras:
 {$$\frac{\rSO(1,1+n)}{\rSO(1+n)}\subset\frac{\rSU(1,1+n)}{\rU(1+n)}
 \subset \frac{\rSO(2,2+n)}{\rSO(2)\times\rSO(2+n)}
 \subset \frac{\rSU(2,2+n)}{\rU(2)\times\rSU(2+n)}$$}

In fact  {$\fsolv({\rSU(1,1+n)}/{\rU(1+n)})$} can be embedded in
more than one way as a subalgebra of
 {$\fs_1=\fsolv({\rSU(2,2+n)}/({\rU(2)\times\rSU(2+n))}$}. For
example,
 {\begin{eqnarray*}&&\fs_2=\rspan\{H_+,Z^{ia},T^{(2,0)}\},\\&&\fs'_2=\rspan\{H_-,Y^{ia},T^{(0,2)}\}\end{eqnarray*}}
are both isomorphic to  {$\fsolv({\rSU(1,1+n)}/{\rU(1+n)})$}, but
they do not commute.  Consequently,
 {$\bigl({\rSU(1,1+n)}/{\rU(1+n)}\bigr)^2$} is not a subgroup of
{${\rSU(2,2+n)}/{(\rU(2)\times\rSU(2+n))}$}.

But there is a generalized contraction  of  {$\fs_1$} in which
 {$\fs_2\oplus\fs'_2$} is a subalgebra. To perform the contraction we
split  {$\fs_1=\fg_0+\fg_{1}+\fg_{2}$}, with {\begin{eqnarray*}
&&\fg_0=\rspan\{H_+,H_-\},\\
&&\fg_{1}=\rspan \{Y^{ia}, Z^{ia}, S_\alpha^{(1,1)}
\},\\&&\fg_{2}=\{T^{(0,2)}, T^{(2,0)},S_\alpha^{(1,-1)}\},
\end{eqnarray*}} and the linear map is
 {$$
\fs_4\rightarrow \fs_4\\
e_0+e_1+e_2\rightarrow e_0+\epsilon e_1+ {\epsilon^2} e_2, \qquad
e_i\in \fg_i. $$} The deformed algebra is {\begin{eqnarray*}
&&[Z^{ia},Z^{jb}]_\e=\epsilon^{ij}\delta^{ab}T^{(2,0)}\\
&&[Y^{ia},Y^{jb}]_\e=\epsilon^{ij}\delta^{ab}T^{(0,2)}\\&&
[Z^{ia},Y^{jb}]_\e=\epsilon \;\delta^{ab}(\delta^{ij}S_2^{(1,1)}+\epsilon^{ij}S_1^{(1,1)})\rightarrow 0\\
&&[Y^{ia},S_1^{(1,-1)}]_\e=\epsilon^2 \; Z^{ia}\rightarrow 0\\
&&[Y^{ia},S_2^{(1,-1)}]_\e= \epsilon^2\; \epsilon^{ij}Z^{ja}\rightarrow 0\\
&&[T^{(0,2)}, S_\alpha^{(1,-1)}]_\e=\epsilon^3\; 2S_\alpha^{(1,1)}\rightarrow 0\\
&&[S_\alpha^{(1,1)},S_\beta^{(1,-1)}]_\e=\epsilon \; \delta_{\alpha\beta}T^{(2,0)}\rightarrow 0\\
&&[H_+, Z^{ia}]_\e=Z^{ia}\\
&&[H_-, Y^{ia}]_\e=Y^{ia}
\end{eqnarray*}}

In the contraction limit, the metric becomes
 {\begin{eqnarray*}&&ds^2=\bigl(d\phi^2+e^{-4\phi} d\tilde t d\tilde
t- 2 e^{-4\phi}y_{1a} d\tilde tdy_{2a} +\\&&
\frac{1}{2}e^{-4\phi}(e^{2\phi}\delta_{ab}+2y_{1a}y_{1b})dy_{2a}
dy_{2b} +
  \\&& \frac{1}{2}e^{-2\phi}dy_{1a}dy_{1a}\bigr)+\\\\&&
\bigl(d\psi^2+e^{-4\psi}dtdt+2e^{-4\psi}z_{2a}dtdz_{1a}+\\&&\frac 12
e^{-4\psi}(e^{2\psi}\delta_{ab}+
2z_{2a}z_{2b})dz_{1a}dz_{1b}+\\&&\frac 12
e^{-2\psi} dz_{2a}dz_{2a}\bigr)+\\\\
 &&+\frac
12e^{-2(\psi+\phi)}d\tilde s_\alpha d\tilde s_\alpha+\frac 12
e^{-2(\psi+\phi)}ds_\alpha ds_\alpha.
\end{eqnarray*}}
The modes  {$s_\alpha, \tilde s_\alpha$} can be exactly set to zero
bu using its field equations, so a trivial truncation of this theory
gives a sigma model on  {$$\frac{\rSU(1,1+n)}{\rU(1+n)}\times
\frac{\rSU(1,1+n)}{\rU(1+n)}$$}

\subsection{Supergravity interpretation}

We consider an {$N=2$} supergravity model coupled to
 {$n+2$}
 hypermultiplets (maximal helicity 1/2) and
{$n+1$} vector multiplets (maximal helicity 1).

We are interested in a particular model which has a ten dimensional
origin. If we consider type IIB SUGRA compactified on the
orientifold  {$T^6/\Z^2$} we obtain an  {$N=4$} theory. When certain
fluxes of the forms in ten dimensions are turned on, this theory has
an  {$N=3$} phase, obtained after the integration of a massive
gravitino multiplet. Turning on other fluxes and performing further
integration we arrive to an  {$N=2$} phase, which is the object of
our interest.

The scalar manifold for this theory is  {$$\cM_Q\times
\cM_{SK}=\frac{\rU(2,2+n)}{\rU(2)\times \rU(2+n)} \times \frac
{\rU(1,1+n)}{\rU(1)\times \rU(1+n)}.$$} (Here  {$n$} refers to the
brane degrees of freedom.) It is the product of a quaternionic
manifold  {$\cM_{Q}$}(hypermultiplets) times a special K\"alher
manifold   {$\cM_{SK}$} (vector multiplets).

We can gauge the translational isometries of the quaternionic
manifold generated by  {$S^{(1,1)}_\alpha$} using two bulk vector
fields. The modes  {$\tilde s_\alpha $} disappear to give mass to
the vectors. The fields {$ s_\alpha $} also acquire a mass through
the potential induced by the gauging. In the large mass limit, these
fields can be set to zero

It can be shown that after the gauging and the integration of the
massive modes, the metric becomes the one of
 {$$\frac{\rSU(1,1+n)}{\rU(1+n)}\times
\frac{\rSU(1,1+n)}{\rU(1+n)}\times \frac{\rSU(1,1+n)}{\rU(1+n)},$$}
where {$\cM_{SK}$} has not been touched \cite{fp}

We see now that the terms set to zero in the metric  by taking the
limit {$\e\rightarrow 0$} are precisely the terms eliminated by the
integration procedure, irrespectively if it is through a Higgs
mechanism or because they acquire mass through the potential.

  The contraction procedure seems a more
general mechanism than the integration. It remains open the
interpretation of other contractions.

  In the case of supersymmetric theories
we are interested in curves in the space of metrics that have two or
more supersymmetric points. These form a smaller set and it is
perhaps
 easier to find a full interpretation of the mechanism.


\section*{Acknowledgments}

We want to thank to our collaborators L. Andrianopoli and  S.
Ferrara.

O. Maci\'a wants to thank the Department of Physics, Theory
Division at CERN for its kind hospitality during the realization
of this work.

This work has been supported by the the Spanish Ministerio de
Educaci\'{o}n y Ciencia through the grant FIS2005-02761 and EU
FEDER funds, by the Generalitat Valenciana, contracts GV04B-226,
GV05/102 and and by the EU network MRTN-CT-2004-005104
'Constituents, Fundamental Forces and Symmetries of the Universe'.


\end{document}